\documentclass[aps,print,twocolumn,hidepacs,hidekeys]{revtex4}%
\usepackage{amssymb}
\usepackage{amsfonts}
\usepackage{amsmath}
\usepackage{graphicx}%
\providecommand{\U}[1]{\protect\rule{.1in}{.1in}}

\begin{document}
\title{Bistable scattering in graphene-coated dielectric nanowires}
\author{Rujiang Li$^{1,2,3}$, Huaping Wang$^{4}$, Bin Zheng$^{1,2,3}$, Shahram Dehdashti$^{1,2,3}$, Erping Li$^{2}$, and Hongsheng Chen$^{1,2,3}$}
\email{hansomchen@zju.edu.cn}
\affiliation{$^{1}$State Key Laboratory of Modern Optical Instrumentation, Zhejiang
University, Hangzhou 310027, China}
\affiliation{$^{2}$College of Information Science and Electronic Engineering, Zhejiang
University, Hangzhou 310027, China}
\affiliation{$^{3}$The Electromagnetics Academy of Zhejiang University, Zhejiang
University, Hangzhou 310027, China}
\affiliation{$^{4}$Institute of Marine Electronics Engineering, Zhejiang University, Hangzhou 310058,
China}

\begin{abstract}
In nonlinear plasmonics, the switching threshold of optical bistability is
limited by the weak nonlinear responses from the conventional Kerr dielectric
media. Considering the giant nonlinear susceptibility of graphene, 
here we develop a nonlinear scattering model under the mean field
approximation and study the bistable scattering in graphene-coated dielectric
nanowires based on the semi-analytical solutions. We find that the switching intensities of bistable scattering 
can be smaller than $1 \text{MW}/\text{cm}^{2}$ at the working frequency.
To further decrease the
switching intensities, we show that the most important factor that
restricts the bistable scattering is the relaxation time of graphene, rather than the chemical potential
and the permittivity of the dielectric nanowire. Our work not only reveals some general
characteristics of graphene-based bistable scattering, but also provides a guidance
to further applications
of optical bistability in the high speed all-optical signal processing.

\end{abstract}
\maketitle



\section{Introduction}

An optical device is said to be bistable if it has two stable output states
under a same input state \cite{Gibbs,RPP45-815}. The output can be switched
between the two stable transmission states, depending upon the history of the
input signal. Optical bistability is a nonlinear phenomenon and it has
promising applications in all-optical switching
\cite{PRE66-055601,APL83-2739,OL33-869}, all-optical logic gates
\cite{IEEEJQE21-377,AO25-1578,OL38-4092,PRB87-195305}, optical transistors
\cite{IEEEJQE17-312,OL28-2506,ncommun4-1778}, and optical memory devices
\cite{PRL61-329,OL30-2575,nphoton4-182,nphoton6-248}. To produce a large
nonlinear effect and decrease the switching threshold of optical bistability, nonlinear media with
high Kerr coefficients are used and sophisticated structures that can
enhance the local field intensities are designed.

Plasmonics is a rapidly growing field which provides an efficient way to
enhance the local electromagnetic fields at subwavelength scales
\cite{nature424-824,nmat9-193}. When nonlinear dielectric media are
incorporated into the plasmonic structures, large nonlinear effects arise due
to the enhanced local fields \cite{nphoton7-737}. Various kinds of nonlinear
plasmonic structures are designed to show the optical bistability
\cite{PRL97-057402,OE15-12368,OE18-13337,PRL108-263905,OE21-13794,APL104-063108,srep6-21741,OE24-22272}%
. However, the commonly used conventional Kerr dielectric materials can only
display very weak nonlinear responses, which restrict further applications
of optical bistability in the high speed all-optical signal processing.

As a newly discovered two dimensional material, graphene shows intriguing
nonlinear characteristics in plasmonics \cite{PR535-101}. Compared with the
conventional Kerr dielectric media, graphene has a giant nonlinear
susceptibility \cite{PRL105-097401}. The use of graphene along with the
enhancement of local plasmonic fields can produce an extremely large nonlinear
effect and consequently decrease the switching threshold of optical
bistability. Recently, some graphene based plasmonic structures have been designed theoretically
or realized experimentally to study the optical bistability, e.g. sandwiched structures
\cite{PRB90-125425,OE23-6479,OM53-80}, a modified Kretschmann-Raether configuration
\cite{srep5-12271}, an Otto configuration \cite{JPD49-255306}, graphene
nanoribbons \cite{PRB92-121407}, and graphene nanobubbles \cite{AOM3-744}.

Compared with those sophisticated structures, graphene-coated dielectric nanowires and nanospheres 
are two kinds of simple structures that are more interested to investigate. 
Such graphene based cylindrical and spherical structures
always demonstrate some typical optical phenomena and can be very useful in lots of applications, such as cloaking
\cite{acsnano5-5855,OE21-12592}, superscattering
\cite{OL40-1651,nanotechnology26-505201,IEEEJSTQE23-4600208}, linear and
nonlinear waveguiding \cite{OL39-5909,OE22-24322,OL38-5244,carbon}, localized
plasmons \cite{PRB91-125414}, and second harmonic generations
\cite{PRB90-035412}. The analytical or semi-analytical solutions calculated
from these simple models are not only useful to reveal some general
characteristics, but also provide a guidance
to further structure design and optimization \cite{srep6-23354}.

In this paper, as a demonstration purpose, we study the bistable scattering in
graphene-coated dielectric nanowires. A nonlinear scattering model is
developed based on the Mie scattering theory. Under the mean field
approximation, this model is solved semi-analytically in both the lossless and lossy cases. 
At the working frequency, the
switching-up and switching-down intensities of bistable scattering are smaller than $1 \text{MW}/\text{cm}^{2}$. 
We show that the most important factor that
restricts the further decrease of switching intensities is the relaxation time of graphene.
Our work reveals some general
characteristics of graphene-based bistable scattering, and provides a theoretical guidance
to further structure design and optimization.


\section{Surface conductivity of graphene}

The optical response of graphene is contributed by both the intraband and
interband transitions. For a doped monolayer graphene, the intraband
transition dominates in the classical frequency range $\hbar\omega\leq\mu_{c}$
\cite{PR535-101}. If the graphene monolayer is placed on the $xy$ plane and a
time-dependent electric field of the form $\mathbf{E}\left(  t\right)  =$
$\hat{x}[E_{x}(\omega)e^{-i\omega t}+c.c.]+\hat{y}[E_{y}(\omega)e^{-i\omega
t}+c.c.]$ is applied, the surface electric current is
\begin{equation}
\mathbf{J}\left(  t\right)  =\hat{x}[j_{x}(\omega)e^{-i\omega t}+c.c.]+\hat
{y}[j_{y}(\omega)e^{-i\omega t}+c.c.], \label{J}%
\end{equation}
where
\begin{align}
j_{i}(\omega)=  &  \sigma_{ii}^{\left(  \omega\right)  }E_{i}+\left(
\sigma_{iijj}^{\left(  3,\omega\right)  }+\sigma_{ijji}^{\left(
3,\omega\right)  }\right)  E_{i}\left\vert E_{j}\right\vert ^{2}\nonumber\\
&  +\sigma_{ijij}^{\left(  3,\omega\right)  }E_{i}^{\ast}E_{j}^{2}%
+\sigma_{iiii}^{\left(  3,\omega\right)  }\left\vert E_{i}\right\vert
^{2}E_{i} \label{j_i}%
\end{align}
is the surface electric current in $i$ direction in the frequency domain,
\begin{equation}
\sigma_{ii}^{\left(  \omega\right)  }=i\frac{e^{2}\mu_{c}}{\pi\hbar^{2}\left(
\omega+i/\tau\right)  } \label{sigma_ii_omega}%
\end{equation}
is the tensor element of the linear surface conductivity,
\begin{equation}
\sigma_{iijj}^{\left(  3,\omega\right)  }=\sigma_{ijji}^{\left(
3,\omega\right)  }=\sigma_{ijij}^{\left(  3,\omega\right)  }=\sigma
_{iiii}^{\left(  3,\omega\right)  }/3=-i\frac{3e^{4}v_{F}^{2}}{8\pi\mu
_{c}\hbar^{2}\omega^{3}} \label{sigma_3,omega}%
\end{equation}
is the tensor element of the nonlinear surface conductivity, $\mu_{c}$ is the
chemical potential, $\tau$ is the relaxation time, $i,j=x,y$, and $c.c.$ denotes the complex conjugate
\cite{PRB90-125425,carbon}. If the applied electric field components satisfy
$E_{x}E_{y}^{\ast}=E_{x}^{\ast}E_{y}$, the surface electric current in $i$
direction in the frequency domain reduces to%
\begin{equation}
j_{i}(\omega)=\sigma_{g}E_{i}(\omega), \label{j_i_1}%
\end{equation}
where%
\begin{equation}
\sigma_{g}=\sigma_{g}^{\left(  1\right)  }+\sigma_{g}^{\left(  3\right)
}\left\vert \mathbf{E}_{\parallel}\right\vert ^{2} \label{sigma_g}%
\end{equation}
is the surface conductivity of graphene,%
\begin{equation}
\sigma_{g}^{\left(  1\right)  }=\sigma_{ii}^{\left(  \omega\right)  }%
=i\frac{e^{2}\mu_{c}}{\pi\hbar^{2}\left(  \omega+i/\tau\right)  },
\label{sigma_g_1}%
\end{equation}
is the linear part,%
\begin{equation}
\sigma_{g}^{\left(  3\right)  }=\sigma_{iiii}^{\left(  3,\omega\right)
}=-i\frac{9e^{4}v_{F}^{2}}{8\pi\mu_{c}\hbar^{2}\omega^{3}}, \label{sigma_g_3}%
\end{equation}
is the coefficient of the nonlinear part, and $\mathbf{E}_{\parallel}=\hat
{x}E_{x}+\hat{y}E_{y}$ is the electric field that is parallel to the graphene
surface. The above calculation results are valid for $\omega\tau\gg1$
\cite{carbon}, where the relaxation time of graphene $\tau$ which describes
the dissipation loss usually ranges from $0.01$ ps to $1$ ps \cite{PR535-101}.

\section{Nonlinear Scattering Model}

\begin{figure}[ptb]
\centering
\vspace{0.3cm}\includegraphics[width=4.5cm]{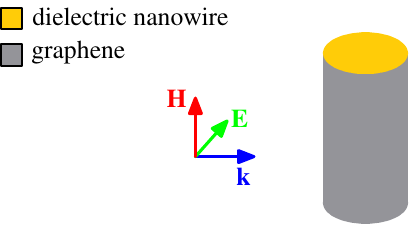} \vspace{-0.0cm}%
\caption{Schematic of the infinitely long graphene-coated
dielectric nanowire. A TM-polarized plane wave with the field components $H_{z}$,
$E_{r}$, and $E_{\theta}$ is incident normally from air onto the structure.
The relative permittivity and permeability of the dielectric nanowire
are $\varepsilon_{r}=2.7$ and $\mu_{r}=1$, respectively, and the radius of the
nanowire is $R=200$ nm.}%
\label{fig1}%
\end{figure}

As shown in Fig. 1, a TM-polarized plane wave with the field components $H_{z}$,
$E_{r}$, and $E_{\theta}$ is incident normally from air onto an infinitely
long graphene-coated dielectric nanowire. The incident magnetic field is
$\mathbf{H}\left(  t\right)  =
\hat{z}\left(H_{0}e^{ik_{0}x-i\omega t}+c.c.\right)/2  ,$
where $k_{0}$ is the wavenumber in free space, $\omega=2\pi
f$ is the angular frequency. The
relative permittivity of the lossless dielectric nanowire is assumed as
$\varepsilon_{r}=2.7$, and the relative permeability is $\mu_{r}=1$. The
radius of the dielectric nanowire is $R=200$ nm. 

Since the thickness of the
graphene coating is extremely small compared with its radius, the graphene coating can
be well characterized as a two dimensional homogenized conducting film, where
the microscopic details are neglected \cite{PRB91-125414}, as shown in Fig. \ref{fig1}. For this structure,
only the azimuthal component of the electric field $E_{\theta}$ contributes to
the nonlinear surface conductivity of graphene, and the surface electric
current along the graphene coating is
\begin{equation}
j_{\theta}(\omega)=\left(  \sigma_{g}^{\left(  1\right)  }+\sigma_{g}^{\left(
3\right)  }\left\vert E_{\theta}\right\vert ^{2}\right)  E_{\theta}(\omega),
\label{j_theta}%
\end{equation}
where $\mathbf{J}\left(  t\right)  =\hat{\theta}( j_{\theta}e^{-i\omega
t}+c.c.) $, $\mathbf{E}\left(  t\right)  =\hat{\theta}( E_{\theta}e^{-i\omega
t}+c.c.) $, and the values of $\sigma_{g}^{\left(  1\right)  }$ and
$\sigma_{g}^{\left(  3\right)  }$ are calculated from Eqs. (\ref{sigma_g_1}%
)-(\ref{sigma_g_3}), respectively.

From Mie scattering theory, the electromagnetic fields inside and outside the dielectric
nanowire can be expressed explicitly \cite{OL40-1651,book1}. According to the
boundary condition of the tangential electric field $E_{\theta}$ at $r=R$, we
have%
\begin{equation}
a_{n}\left[  J_{n}^{\prime}\left(  k_{0}R\right)  +s_{n}H_{n}^{\left(
1\right)  \prime}\left(  k_{0}R\right)  \right]  =\frac{1}{\sqrt
{\varepsilon_{r}}}b_{n}J_{n}^{\prime}\left(  kR\right)  ,
\label{boundary condition 1}%
\end{equation}
where $a_{n}=\delta_{n}i^{n}$ ($\delta_{n}=1$ for $n=0$ and $\delta_{n}=2$ for
$n\neq0$), $k=k_{0}\sqrt{\varepsilon_{r}}$ is the wavenumber in the dielectric
nanowire, $s_{n}$ is the scattering coefficient, and $J_{n}$ and
$H_{n}^{\left(  1\right)  }$ are the $n$-th order Bessel function of the first
kind and Hankel function of the first kind, respectively. Then according to
the another boundary condition from $H_{z}$, we have%
\begin{align}
&  \sum\limits_{n=0}^{\infty}a_{n}\left[  J_{n}\left(  k_{0}R\right)
+s_{n}H_{n}^{\left(  1\right)  }\left(  k_{0}R\right)  \right]  \cos\left(
n\theta\right) \nonumber\\
=  &  \sum\limits_{n=0}^{\infty}b_{n}\left[  J_{n}\left(  kR\right)
+i\frac{\sigma_{g}\eta_{0}}{\sqrt{\varepsilon_{r}}}J_{n}^{\prime}\left(
kR\right)  \right]  \cos\left(  n\theta\right)  , \label{boundary condition 2}%
\end{align}
where%
\begin{equation}
\sigma_{g}=\sigma_{g}^{\left(  1\right)  }+\frac{\sigma_{g}^{\left(  3\right)
}\eta_{0}^{2}H_{0}^{2}}{4\varepsilon_{r}}\left\vert \sum\limits_{m=0}^{\infty
}b_{m}J_{m}^{\prime}\left(  kR\right)  \cos\left(  m\theta\right)  \right\vert
^{2}. \label{sigma_g_1_1}%
\end{equation}
From this general boundary condition, we can see that scattering terms with different
orders are coupled via the nonlinear surface conductivity of graphene. 
Since a scattering term with a given order contributes to the scattering terms with
other orders, the total surface conductivity of graphene is inhomogeneous along the azimuthal
direction. 

In Eq. (\ref{j_theta}), the nonlinear part is usually small compared with the linear part. We can use
the mean field approximation to simplify the nonlinear scattering model, where
the field intensity at the circumference of the graphene coating is represented by
the average value \cite{PRL108-263905,srep6-21741}. From Eq. (\ref{sigma_g_1_1}), we have
\begin{align}
\sigma_{g}=  &  \sigma_{g}^{\left(  1\right)  }+\frac{\sigma_{g}^{\left(
3\right)  }\eta_{0}^{2}H_{0}^{2}}{8\pi\varepsilon_{r}}\sum\limits_{m=0}%
^{\infty}\sum\limits_{m^{\prime}=0}^{\infty}b_{m}b_{m^{\prime}}^{\ast}%
J_{m}^{\prime}\left(  kR\right)  J_{m^{\prime}}^{\prime\ast}\left(  kR\right)
\nonumber\\
&  \times\int\nolimits_{0}^{2\pi}\cos\left(  m\theta\right)  \cos\left(
m^{\prime}\theta\right)  d\theta, \label{sigma_g_2}%
\end{align}
which simplifies to%
\begin{align}
\sigma_{g} =  &  \sigma_{g}^{\left(  1\right)  }+\frac{\sigma_{g}^{\left(
3\right)  }\eta_{0}^{2}H_{0}^{2}}{4\varepsilon_{r}}\left\vert b_{0}\right\vert
^{2}J_{0}^{\prime2}\left(  kR\right) \nonumber\\
&  +\frac{\sigma_{g}^{\left(  3\right)  }\eta_{0}^{2}H_{0}^{2}}{8\varepsilon
_{r}}\sum\limits_{m=1}^{\infty}\left\vert b_{m}\right\vert ^{2}J_{m}^{\prime
2}\left(  kR\right)  . \label{sigma_g_4}%
\end{align}
Thus, under the mean field approximation, the surface conductivity of graphene
is independent on the $\theta$. But scattering terms with different orders are
still coupled.

If the radius of the dielectric nanowire is far smaller than the
wavelength, the total scattering is dominated by the $m=1$ term
\cite{OL40-1651} (Supplement 1, Section 1). The surface conductivity of graphene reduces to%
\begin{equation}
\sigma_{g}=\sigma_{g}^{\left(  1\right)  }+\frac{\sigma_{g}^{\left(  3\right)
}\eta_{0}^{2}H_{0}^{2}}{8\varepsilon_{r}}\left\vert b_{1}\right\vert ^{2}%
J_{1}^{\prime2}\left(  kR\right)  . \label{sigma_g_5}%
\end{equation}
In this case, the surface conductivity of graphene only depends on the
first order scattering term, and the other scattering terms are coupled
with the first order scattering term via the surface conductivity. However,
the nonlinear part of the surface conductivity is usually small compared with the
linear part, and the contribution from the first order scattering term to the
other terms are small. The total scattering can be simplified further by
neglecting the weak coupling between the terms. For $n=1$, Eq.
(\ref{boundary condition 2}) reduces to%
\begin{align}
&  \sqrt{\varepsilon_{r}}a_{1}\left[  J_{1}\left(  k_{0}R\right)  +s_{1}%
H_{1}^{\left(  1\right)  }\left(  k_{0}R\right)  \right] \nonumber\\
&  -b_{1}\left[  \sqrt{\varepsilon_{r}}J_{1}\left(  kR\right)  +i\sigma
_{g}^{\left(  1\right)  }\eta_{0}J_{1}^{\prime}\left(  kR\right)  \right]
\nonumber\\
=  &  \alpha b_{1}\left\vert b_{1}\right\vert ^{2}J_{1}^{\prime3}\left(
kR\right)  , \label{boundary condition 2_1}%
\end{align}
where $\alpha=i\sigma_{g}^{\left(  3\right)  }\eta_{0}^{3}H_{0}^{2}/\left(
8\varepsilon_{r}\right)  $. This is the boundary condition for $H_{z}$ under
the mean field approximation.

From the two boundary conditions shown by Eqs.
(\ref{boundary condition 1}) and (\ref{boundary condition 2_1}), we obtain a
cubic nonlinear equation for the first order scattering coefficient $s_{1}$ (Supplement 1, Section 2).
In order to solve this equation, for simplicity, we can use
the approximations of $k_{0}R\ll1$ and $kR\ll1$ since the wavelengths are much
larger than the radius of the dielectric nanowire. Using the asymptotic
expansions of the Bessel function and Hankel function \cite{book2},
the nonlinear equation reduces to
\begin{align}
&  c^{\left(  1\right)  }+c_{1}^{\left(  1\right)  }s_{1}+c_{1^{\ast}%
}^{\left(  1\right)  }s_{1}^{\ast}+c_{11}^{\left(  1\right)  }s_{1}%
^{2}+c_{11^{\ast}}^{\left(  1\right)  }\left\vert s_{1}\right\vert
^{2}\nonumber\\
&  +c_{11^{\ast}1}^{\left(  1\right)  }\left\vert s_{1}\right\vert ^{2}%
s_{1}\nonumber\\
=  &  0, \label{nonlinear equation_1}%
\end{align}
where the coefficients are%
\begin{align}
c^{\left(  1\right)  }=  &  \frac{1}{4}\alpha\varepsilon_{r}+\frac{1}%
{4}\left[  \left(  \varepsilon_{r}-1\right)  k_{0}R+i\sigma_{g}^{\left(
1\right)  }\eta_{0}\right]  ,\label{c1_w}\\
c_{1}^{\left(  1\right)  }=  &  i\frac{2\alpha\varepsilon_{r}}{\pi k_{0}%
^{2}R^{2}} +i\frac{1}{\pi k_{0}^{2}R^{2}}\left[  \left(  \varepsilon
_{r}+1\right)  k_{0}R+i\sigma_{g}^{\left(  1\right)  }\eta_{0}\right]
,\label{c1_w_1}\\
c_{1^{\ast}}^{\left(  1\right)  }=  &  -i\frac{\alpha\varepsilon_{r}}{\pi
k_{0}^{2}R^{2}},\label{c1_w_1s}\\
c_{11}^{\left(  1\right)  }=  &  -\frac{4\alpha\varepsilon_{r}}{\pi^{2}%
k_{0}^{4}R^{4}},\label{c1_w_11}\\
c_{11^{\ast}}^{\left(  1\right)  }=  &  \frac{8\alpha\varepsilon_{r}}{\pi
^{2}k_{0}^{4}R^{4}},\label{c1_w_11s}\\
c_{11^{\ast}1}^{\left(  1\right)  }=  &  i\frac{16\alpha\varepsilon_{r}}%
{\pi^{3}k_{0}^{6}R^{6}}. \label{c1_w_11s1}%
\end{align}
Specially, if the nonlinear surface conductivity
of graphene is neglected, namely in the limit of $H_{0}=0$, the scattering coefficient reduces to%
\begin{equation}
s_{1}=i\frac{\pi k_{0}^{2}R^{2}}{4}\frac{\left(  \varepsilon_{r}-1\right)
k_{0}R+i\sigma_{g}^{\left(  1\right)  }\eta_{0}}{\left(  \varepsilon
_{r}+1\right)  k_{0}R+i\sigma_{g}^{\left(  1\right)  }\eta_{0}}.
\label{s_1_linear}%
\end{equation}

After calculating the scattering coefficient, the normalized scattering cross
section section (NSCS) can be evaluated as%
\begin{equation}
\text{NSCS}=2\left\vert s_{1}\right\vert ^{2}, \label{NSCS}%
\end{equation}
where only the first order scattering term is considered.

\section{Bistable Scattering}

In the following, we study the bistable scattering in graphene-coated dielectric nanowires based on the nonlinear scattering model.
As an example, the chemical potential of graphene is chosen as
$\mu_{c}=0.35$ eV.

According to Eqs. (\ref{sigma_g})-(\ref{sigma_g_3}), the surface conductivity of graphene has two parts, where the linear part is a complex number
and the nonlinear part is a pure imaginary number. From Eqs. (\ref{c1_w})-(\ref{c1_w_11s1}),
only the two coefficients $c^{\left(  1 \right)  }$ and $c^{\left(  1 \right)  }_{1}$ are complex, while
the other coefficients are all real. Thus, Eq. (\ref{nonlinear equation_1}) is a nonlinear equation with
complex coefficients and it only has complex solutions generally. Specially, if the relaxation time
of graphene is infinite, namely if there is no dissipation loss, the linear surface conductivity in Eq. (\ref{sigma_g_1}) only has the
imaginary part and Eq.
(\ref{nonlinear equation_1}) has pure imaginary solutions.

\subsection{Lossless Case}

First we consider the lossless case, where the relaxation time
of graphene is infinite. Since $s_{1}$ reduces to a pure imaginary number, we
define $s_{1}=is_{01}$. Then Eq. (\ref{nonlinear equation_1}) reduces to%
\begin{equation}
c^{\left(  1\right)  }+c_{1}^{\left(  1\right)  }s_{01}+c_{11}^{\left(
1\right)  }s_{01}^{2}+c_{111}^{\left(  1\right)  }s_{01}^{3}=0,
\label{nonlinear equation_2}%
\end{equation}
where the coefficients are
\begin{align}
c^{\left(  1\right)  }=  &  -\frac{\sigma_{g,i}^{\left(  3\right)  }\eta
_{0}^{2}I_{0}}{8}+\frac{1}{2}\left[  \left(  \varepsilon_{r}-1\right)
k_{0}R-\sigma_{g,i}^{\left(  1\right)  }\eta_{0}\right]  ,\label{c1_w1}\\
c_{1}^{\left(  1\right)  }=  &  \frac{3\sigma_{g,i}^{\left(  3\right)  }%
\eta_{0}^{2}I_{0}}{2\pi k_{0}^{2}R^{2}}-\frac{2}{\pi k_{0}^{2}R^{2}}\left[
\left(  \varepsilon_{r}+1\right)  k_{0}R-\sigma_{g,i}^{\left(  1\right)  }%
\eta_{0}\right]  ,\label{c1_w1_1}\\
c_{11}^{\left(  1\right)  }=  &  -\frac{6\sigma_{g,i}^{\left(  3\right)  }%
\eta_{0}^{2}I_{0}}{\pi^{2}k_{0}^{4}R^{4}},\label{c1_w1_11}\\
c_{111}^{\left(  1\right)  }=  &  \frac{8\sigma_{g,i}^{\left(  3\right)  }%
\eta_{0}^{2}I_{0}}{\pi^{3}k_{0}^{6}R^{6}}, \label{c1_w1_111}%
\end{align}
the linear and nonlinear surface conductivities are%
\begin{align}
\sigma_{g,i}^{\left(  1\right)  }  &  =\operatorname{Im}\sigma_{g}^{\left(
1\right)  }=\frac{e^{2}\mu_{c}}{\pi\hbar^{2}\omega},\label{sigma_gi_1}\\
\sigma_{g,i}^{\left(  3\right)  }  &  =\operatorname{Im}\sigma_{g}^{\left(
3\right)  }=-\frac{9e^{4}v_{F}^{2}}{8\pi\mu_{c}\hbar^{2}\omega^{3}},
\label{sigma_gi_3}%
\end{align}
and $I_{0}=\eta_{0}H_{0}^{2}/2$ is the incident field intensity. Specially, in
the limit of $I_{0}=0$, the nonlinear equation reduces to the linear case, and
the scattering coefficient is
\begin{equation}
s_{1}=i s_{01}  =i\frac{\pi k_{0}^{2}R^{2}}{4}\frac{\left(  \varepsilon
_{r}-1\right)  k_{0}R-\sigma_{g,i}^{\left(  1\right)  }\eta_{0}}{\left(
\varepsilon_{r}+1\right)  k_{0}R-\sigma_{g,i}^{\left(  1\right)  }\eta_{0}%
}.\label{s01_linear}
\end{equation}

\begin{figure}[ptb]
\centering
\vspace{0.0cm} \includegraphics[width=8.4cm]{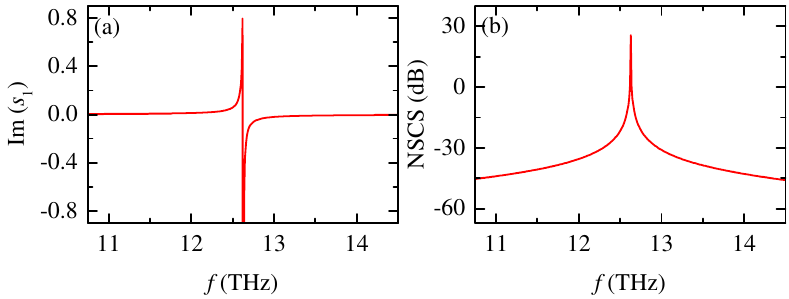} \vspace{-0.0cm}%
\caption{(a) The scattering coefficients and (b) NSCSs at
different frequencies, where $I_{0}=0$, $\mu_{c}=0.35$ eV, and $\tau = \infty$.}%
\label{fig2}%
\end{figure}

The scattering coefficient in the linear case can be directly calculated from
Eq. (\ref{s01_linear}). Figs. \ref{fig2}(a)-(b) show the scattering coefficients
and NSCSs at different frequencies, where $I_{0}=0$, $\mu_{c}=0.35$ eV, and $\tau = \infty$. The NSCS exhibits the superscattering at
$f_{0}=12.63$ THz, where $\left(  \varepsilon_{r}+1\right)  k_{0}%
R=\sigma_{g,i}^{\left(  1\right)  }\eta_{0}$, and the scattering coefficient
shows a phase change of $\pi$ at the superscattering frequency \cite{OL40-1651}. According to
Fig. \ref{fig2}(b), in order to enhance the local field intensity along the
graphene coating in the nonlinear case, the working frequency for bistable scattering 
should be near to the superscattering frequency. 

Eq. (\ref{nonlinear equation_2}) can be solved analytically \cite{book2}, where only the real solutions
are physically acceptable. When $I_{0}\neq
0$, we define%
\begin{align}
p =  &  \frac{c_{1}^{\left(  1\right)  }}{c_{111}^{\left(  1\right)  }}%
-\frac{c_{11}^{\left(  1\right)  2}}{3c_{111}^{\left(  1\right)  2}}%
,\label{p}\\
q =  &  \frac{c^{\left(  1\right)  }}{c_{111}^{\left(  1\right)  }}%
+\frac{2c_{11}^{\left(  1\right)  3}}{27c_{111}^{\left(  1\right)  3}}%
-\frac{c_{11}^{\left(  1\right)  }c_{1}^{\left(  1\right)  }}{3c_{111}%
^{\left(  1\right)  2}}, \label{q}%
\end{align}
then the three solutions can be written as%
\begin{align}
s_{01,1} =  &  A^{1/3}+B^{1/3}+\Gamma,\label{s01,1}\\
s_{01,2} =  &  \omega_{+}A^{1/3}+\omega_{-}B^{1/3}+\Gamma,\label{s01,2}\\
s_{01,3} =  &  \omega_{-}A^{1/3}+\omega_{+}B^{1/3}+\Gamma, \label{s01,3}%
\end{align}
where $A=-q/2+\sqrt{\Delta}$, $B=-q/2-\sqrt{\Delta}$, $\Gamma=-b/\left(
3a\right)  $, $\omega_{\pm}=\left(  -1\pm\sqrt{3}i\right)  /2$, and
\begin{equation}
\Delta=\left(  \frac{q}{2}\right)  ^{2}+\left(  \frac{p}{3}\right)  ^{3}.
\label{delta}%
\end{equation}
If $\Delta>0$, $A$ and $B$ are both real numbers with $A\neq B$. The three
solutions are%
\begin{align}
s_{01,1} =  &  A^{1/3}+B^{1/3}+\Gamma,\label{s01,1_1}\\
s_{01,2} =  &  -\frac{1}{2}\left(  A^{1/3}+B^{1/3}\right)  +\frac{\sqrt{3}}%
{2}i\left(  A^{1/3}-B^{1/3}\right)  +\Gamma,\label{s01,2_1}\\
s_{01,3} =  &  -\frac{1}{2}\left(  A^{1/3}+B^{1/3}\right)  -\frac{\sqrt{3}}%
{2}i\left(  A^{1/3}-B^{1/3}\right)  +\Gamma, \label{s01,3_1}%
\end{align}
namely, only $s_{01,1}$ is the real solution, and $s_{01,2}$ and $s_{01,3}$
are conjugated complex solutions. If $\Delta=0$, $A$ and $B$ are both real
numbers with $A=B$. The three solutions are%
\begin{align}
s_{01,1} =  &  2A^{1/3}+\Gamma,\label{s01,1_2}\\
s_{01,2} =  &  -A^{1/3}+\Gamma,\label{s01,2_2}\\
s_{01,3} =  &  -A^{1/3}+\Gamma, \label{s01,3_2}%
\end{align}
namely the three solutions are all real and $s_{01,2}=s_{01,3}$. If $\Delta
<0$, $A$ and $B$ are conjugated complex numbers with $A^{1/3}=C+iD$ and
$B^{1/3}=C-iD$. The three solutions are%
\begin{align}
s_{01,1} =  &  2C+\Gamma,\label{s01,1_3}\\
s_{01,2} =  &  -C-\sqrt{3}D+\Gamma,\label{s01,2_3}\\
s_{01,3} =  &  -C+\sqrt{3}D+\Gamma, \label{s01,3_3}%
\end{align}
namely the three solutions are all real solutions.

Thus, $\Delta$ is the discriminant of the cubic nonlinear equation. According
to Eq. (\ref{delta}), we have%
\begin{equation}
\frac{\Delta}{\Delta_{0}}=-4\left[  \left(  \varepsilon_{r}+1\right)
k_{0}R-\sigma_{g,i}^{\left(  1\right)  }\eta_{0}\right]  ^{3}+27k_{0}^{2}%
R^{2}\sigma_{g,i}^{\left(  3\right)  }\eta_{0}^{2}I_{0}, \label{delta_1}%
\end{equation}
where $\Delta_{0}=\pi^{6}k_{0}^{12}R^{12}/\left(  6912\sigma_{g,i}^{\left(
3\right)  3}\eta_{0}^{6}I_{0}^{3}\right)  $ is always a negative number. If the working
frequency of bistable scattering is chosen as $f\geq f_{0}$, the first term in Eq. (\ref{delta_1}) is
negative and the determinant $\Delta$ is always positive. Thus Eq.
(\ref{nonlinear equation_2}) has only one real solution. Whereas, if $f<f_{0}%
$, the first term in Eq. (\ref{delta_1}) is positive and the determinate
$\Delta$ may be positive or negative depending upon the value of $I_{0}$.
Thus Eq. (\ref{nonlinear equation_2}) has three real solutions for%
\begin{equation}
0<I_{0}<\frac{4\left[  \left(  \varepsilon_{r}+1\right)  k_{0}R-\sigma
_{g,i}^{\left(  1\right)  }\eta_{0}\right]  ^{3}}{27k_{0}^{2}R^{2}\sigma
_{g,i}^{\left(  3\right)  }\eta_{0}^{2}}, \label{I0_1}%
\end{equation}
two different real solutions for
\begin{equation}
I_{0}=\frac{4\left[  \left(  \varepsilon_{r}+1\right)  k_{0}R-\sigma
_{g,i}^{\left(  1\right)  }\eta_{0}\right]  ^{3}}{27k_{0}^{2}R^{2}\sigma
_{g,i}^{\left(  3\right)  }\eta_{0}^{2}}, \label{I0_2}%
\end{equation}
and only one real solution for $I_{0}=0$ and%
\begin{equation}
I_{0}>\frac{4\left[  \left(  \varepsilon_{r}+1\right)  k_{0}R-\sigma
_{g,i}^{\left(  1\right)  }\eta_{0}\right]  ^{3}}{27k_{0}^{2}R^{2}\sigma
_{g,i}^{\left(  3\right)  }\eta_{0}^{2}}. \label{I0_3}%
\end{equation}

\begin{figure}[ptb]
\centering
\vspace{0.0cm} \includegraphics[width=8.4cm]{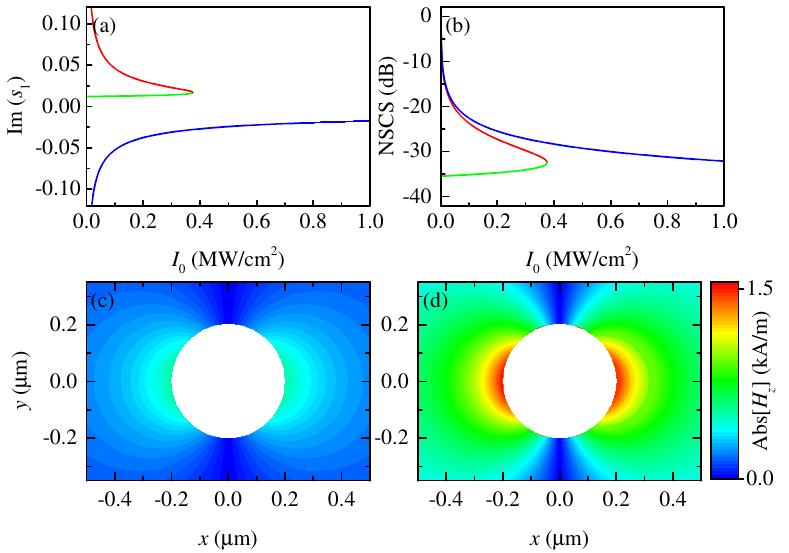} \vspace{-0.0cm}%
\caption{(a) The scattering coefficients and (b) NSCSs for
different incident field intensities. The curves with different colors correspond to different branches.
(c)-(d) For $I_{0}=0.2$  $\text{MW}/\text{cm}^{2}$, the 
scattered fields which corresponds to the ``off'' and ``on'' states are shown,
respectively, where the fields inside the dielectric nanowire are not shown. The parameters are $f=12.00$ THz, $\mu_{c}=0.35$ eV,
and $\tau = \infty$.}%
\label{fig3}%
\end{figure}

According to the above discussion, the working frequency of bistable scattering
should be smaller than the superscattering frequency (Supplement 1, Section 3). 
Figs. \ref{fig3}(a)-(b) show the scattering coefficients and NSCSs for different
incident field intensities at $f=12.00$ THz, where $\mu_{c}=0.35$ eV and $\tau = \infty$. 
Bistable scattering occurs under these parameters. 
For $I_{0}=0$, Eq. (\ref{nonlinear equation_2}) has only one real solution, which
corresponds to the starting points of the green curves. If $I_{0}$ increases
from 0, the second term in Eq. (\ref{delta_1}) increases. According to Eq.
(\ref{I0_1}), if the incident field intensity is smaller than a threshold, one
incident field intensity corresponds to three scattering coefficients, where
two coefficients are positive and one coefficient is negative. Specially, if
$I_{0}$ increases to the threshold where $\Delta=0$, the two positive scattering coefficients are
equal. If $I_{0}$ continue increases, only the negative scattering coefficient still exists
according to Eq. (\ref{I0_3}).

The fulfillment of $\Delta<0$ at a certain range of incident field intensity
insures the existence of bistable scattering. As shown in Fig. \ref{fig3}(b),
the bistable behavior can be used as a scattering switch. The bottom green
curve and the top blue curve are two stable scattering states, and the middle red
curve is an instable scattering state. Due to the instability, usually the scattering
cannot exist on this state. The top and bottom stable states can be treated as
``on'' and ``off'' states of a scattering switch, respectively. For $I_{0}=0.2$  $\text{MW}/\text{cm}^{2}$, the 
scattered fields which corresponds to the ``off'' and ``on'' states are shown in Fig. \ref{fig3}(c)-(d),
respectively, where the fields in the dielectric nanowire are not shown. The scattering switch has a high contrast between
the fields. According to the
analytical results, there are two intensity thresholds to control the switch. The scattering changes from ``off'' to ``on''
at the switching-up intensity $T_{\text{on}}$, and the scattering
changes from ``on'' to ``off'' at the switching-down intensity $T_{\text{off}}$. In this lossless
nonlinear scattering model, $T_{\text{off}}=0$ and
\begin{equation}
T_{\text{on}}=\frac{4\left[  \left(  \varepsilon_{r}+1\right)  k_{0}%
R-\sigma_{g,i}^{\left(  1\right)  }\eta_{0}\right]  ^{3}}{27k_{0}^{2}%
R^{2}\sigma_{g,i}^{\left(  3\right)  }\eta_{0}^{2}}. \label{T_on}%
\end{equation}
The switching-up intensity $T_{\text{on}}$ is inversely dependent on the nonlinear surface
conductivity of graphene, namely a low intensity threshold requires a large $\sigma
_{g,i}^{\left(  3 \right)  }$.

\subsection{Lossy Case}

\begin{figure}[ptb]
\centering
\vspace{0.0cm} \includegraphics[width=8.5cm]{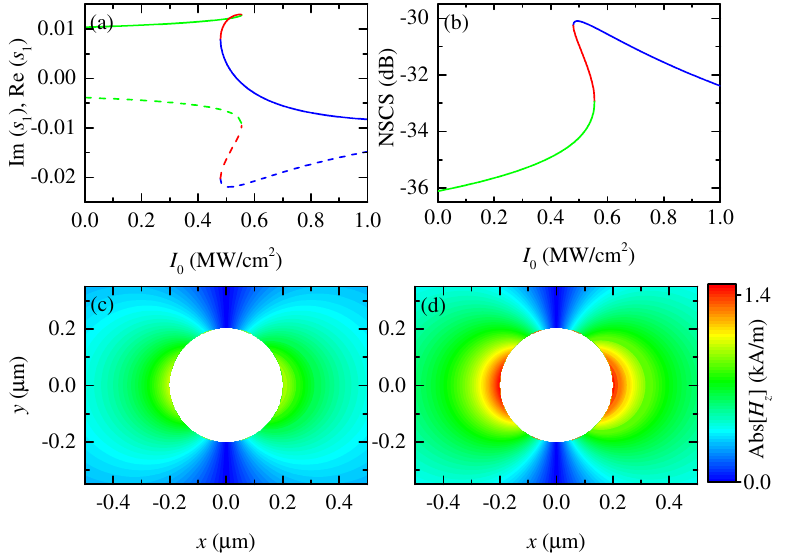} \vspace{-0.0cm}%
\caption{(a) The scattering coefficients and (b) NSCSs for
different incident field intensities. The curves with different colors correspond to different branches.
In (a), the solid curve denotes $\text{Im}\left( s_{1}\right)$,
and the dashed curve denotes $\text{Re}\left( s_{1}\right)$. (c)-(d) For $I_{0}=0.5$  $\text{MW}/\text{cm}^{2}$, the 
scattered fields which corresponds to the ``off'' and ``on'' states are shown,
respectively, where the fields inside the dielectric nanowire are not shown. The parameters are $f=12.00$ THz, $\mu_{c}=0.35$ eV,
and $\tau=0.30$ ps.}%
\label{fig4}%
\end{figure}

Next, we consider the lossy case, where the relaxation time of graphene is finite. 
In this case, Eq. (\ref{nonlinear equation_1}) can be solved numerically (Supplement 1, Section 3).
Figs. \ref{fig4}(a)-(b) show the scattering coefficients and NSCSs for different
incident field intensities at $f=12.00$ THz, where $\mu_{c}=0.35$ eV and $\tau=0.30$ ps.
At this working frequency, both
the scattering coefficient and NSCS show more interesting phenomena.

In Fig. \ref{fig3}(a) where $\tau = \infty$, the solid red and blue curves extend to the positive and negative infinities,
respectively. In contrast, the solid red and blue curves in Fig. \ref{fig4}(a)
are connected to each other at a certain incident field intensity, and a loop is formed by the three curves. 
Meanwhile, due to the finite relaxation time of graphene, the real part of the scattering coefficient is nonzero,
and it demonstrates a bistable behavior with a dip,
as shown by the dashed curves in Fig. \ref{fig4}(a).

\begin{figure*}[ptb]
\centering
\vspace{0.0cm} \includegraphics[width=13.4cm]{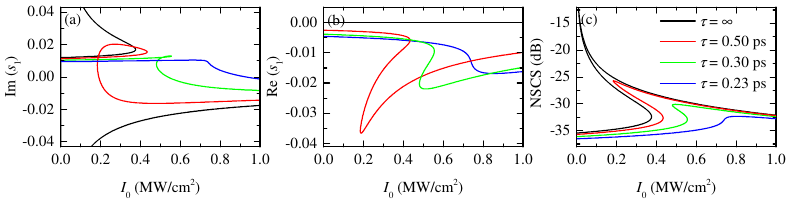} \vspace{-0.0cm}%
\caption{(a) The imaginary and (b) real parts of the scattering
coefficients and (c) the NSCSs for different relaxation times of graphene, where $f=12.00$ THz,
$\mu_{c}=0.35$, and the values of $\tau$ are
taken as $0.23$ ps, $0.30$ ps, and $0.50$ ps, respectively. The cases for $\tau = \infty$ are also plotted for comparison}%
\label{fig5}%
\end{figure*}

Both the real and imaginary parts of the scattering coefficient contribute to
the NSCS. Compared with the NSCS in Fig. \ref{fig3}(b), the NSCS in Fig. \ref{fig4}(b)
has larger switching-up and switching-down intensities.
Specially, the switching-down intensity is no longer equal to zero and the difference
between the switching intensities defined as $\Delta T= T_{\text{on}}-T_{\text{off}}$ is decreased. 
Similarly, we also plot the 
scattered fields which corresponds to the ``off'' and ``on'' states in Fig. \ref{fig4}(c)-(d),
respectively, where $I_{0}=0.5$  $\text{MW}/\text{cm}^{2}$, and the fields in the dielectric nanowire are not shown. 
The contrast between the fields is decreased due to the finite relaxation time of graphene.
This implies that, the relaxation time of graphene
plays a significant role in the bistable scattering.

Finally, we would like to note the validity our nonlinear scattering model. In Figs. \ref{fig3}-\ref{fig4}, the ratio between the nonlinear 
and linear parts of the surface conductivity of graphene is less than $1 \times 10^{-10}$. This implies that the
$\theta$ dependent nonlinear part only adds a small perturbation to the total surface conductivity and the field intensity along the
graphene coating can be homogenized using the mean field approximation. 
Besides, in both the lossless
and lossy cases, we use the approximations of $k_{0}R\ll1$ and $kR\ll1$, where the Bessel function and Hankel function are replaced by their respective asymptotic expansions in the nonlinear scattering model. By comparing the NSCSs with and without these approximations,
we show that this kind of approximations is also valid (Supplement 1, Section 4).  
Our nonlinear scattering model provides a simple way to understand the bistable scattering.

\section{Discussion}

According to the discussion in the previous section, the bistable scattering strongly dependents on the relaxation time of graphene,
which describes the dissipation loss. Figs. \ref{fig5}(a)-(c) show the
imaginary and real parts of the scattering coefficients and the NSCSs for different
relaxation times, where $f=12.00$ THz, $\mu_{c}=0.35$ eV, and the values of $\tau$ are taken as
$0.23$ ps, $0.30$ ps, and $0.50$ ps, respectively. The cases for $\tau = \infty$ are also plotted for comparison. 

For $\tau= \infty$, two branches of the imaginary part of $s_{1}$
extend to the positive infinity and negative infinity, respectively. If $\tau$
decreases, the two branches connect to each other when the incident field
intensity $I_{0}=T_{\text{off}}$, as shown in Fig. \ref{fig5}(a). A loop is
formed by the three branches due to the existence of the switching-down intensity $T_{\text{off}}$ and
the switching-up intensity $T_{\text{on}}$. However, if $\tau$ continues to decrease, the loop shrinks gradually.
When the relaxation time is small enough, the loop disappears and there is no bistable behavior.

The real part of the scattering coefficient is emerged due to the finite relaxation
time of graphene. For $\tau=\infty$, the real part of $s_{1}$ is always equal to
zero. If $\tau$ decreases, two branches of the real part form a dip, as shown in Fig. \ref{fig5}(b).
The three branches are connected to each other, and exhibit a bistable behavior. 
However, if $\tau$ continues to decrease, the dip becomes smooth gradually.
When the relaxation time is small enough, the dip almost disappears and there is no bistable behavior.

The simultaneous existences of the loop from the imaginary part and the dip
from the real part of the scattering coefficient indicate the existence of
the bistable scattering. As shown in Fig. \ref{fig5}(c), for different relaxation
times of graphene, the NSCSs show bistable scattering with different
hysteresis curves. With the decreasing of $\tau$, the switching-up and switching-down intensities
both decrease, and the difference between the switching intensities also decreases. 
Meanwhile, the maximum NSCS at the curve
decreases sharply. If $\tau$ continues to decrease, the bistable scattering
disappears and the hysteresis curve becomes a smooth curve
gradually. 

From Fig. \ref{fig5}, we can see that, if the incident frequency is fixed,
in order to show the bistable scattering, the relaxation time of graphene should be large enough.
From one hand, if the relaxation time is small, we need large switching-up and switching-down intensities,
and the bistable scattering requires a large switching threshold.
From another hand, if the relaxation time of graphene is too small, there is no bistable scattering at all.
Thus, the working frequency and the performance of bistable scattering are restricted by the dissipation
loss of graphene.   

\begin{figure}[ptb]
\centering
\vspace{0.0cm} \includegraphics[width=8.4cm]{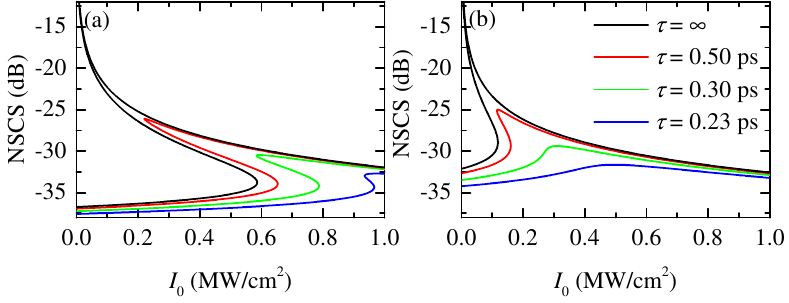} \vspace{-0.0cm}%
\caption{The NSCSs for different relaxation times of graphene,
where $\mu_{c}=0.35$ eV, $f=11.90$ THz in (a) and $f=12.20$ THz in (b),
and the values of $\tau$ are taken as $0.23$ ps, $0.30$
ps, and $0.50$ ps, respectively. The cases for $\tau = \infty$ are also plotted for comparison.}%
\label{fig6}%
\end{figure}

In order to further explain this restriction, in Fig. \ref{fig6} we show the bistable scatterings at $f=11.90$ THz and $f=12.20$ THz, respectively. 
At $f=11.90$ THz, bistable scatterings are shown for all the three values of $\tau$.
While at $f=12.20$ THz, bistable scattering is only shown for $\tau = 0.50$ ps. 
In other words, at a low working frequency, bistable scattering occurs for a small relaxation time, and
a large switching threshold is required. To decrease the switching threshold, the working frequency
should be increased towards to the superscattering frequency. But if the working frequency is high,
the bistable scattering is more sensitive to the relaxation time. 
To realize the bistable scattering at a high working frequency, the relaxation time of
graphene must be large enough.

The chemical potential of graphene and the permittivity of the dielectric nanowire also affect
the bistable scattering (Supplement 1, Section 5). According to Ref. \cite{OL40-1651}, the changes of the chemical potential
and the permittivity are equivalent with the change of the superscattering frequency.
And the shift of the superscattering frequency is equivalent with the reverse shift of the working frequency.
The performance of bistable scattering is still restricted by the relaxation time of graphene.  

\section{Conclusion}

In conclusion, we have developed a nonlinear scattering model for a graphene-coated dielectric nanowire
under the mean field approximation. The bistable scattering is discussed based on the semi-analytical solutions.
We have found that the relaxation time of graphene is the main restriction to further decrease the switching threshold
of bistable scattering, although the other parameters including the working frequency, the chemical potential of graphene,
and the permittivity of the dielectric nanowire can be tuned freely. The development of graphene–boron nitride heterostructures
that can support low-loss plasmons may shed new light on optical bistability in nonlinear graphene plasmonics \cite{nnano5-722,nmat14-421}.  
Our work provides a theoretical guidance to further structure design and optimization, 
and application of optical bistability in the high speed all-optical signal processing.

\section*{Funding Information}

National Natural Science Foundation of China (61625502, 61574127, 61601408, 61550110245);
ZJNSF (LY17F010008); Fundamental Research Funds for the
Central Universities (2016QNA5006); Top-Notch Young Talents
Program of China; Innovation Joint Research Center for
Cyber-Physical-Society System.





\bigskip \noindent See \href{Optica-suppl-materials-template}{Supplement 1} for supporting content.

\end{document}